\documentclass[preprint,superscriptaddress, nofootinbib, aps,pra]{revtex4}
\usepackage{braket}             
\usepackage{graphicx}           
\usepackage{bm}                 
\usepackage{amsmath,amssymb,amsthm}
\usepackage[usenames, dvipsnames]{color}
\usepackage{xcolor}
\usepackage{hyperref}           

\usepackage{mathtools}
\usepackage{soul}               
\usepackage{mathrsfs}
\usepackage{subfigure}
\usepackage{ulem}
\usepackage{colonequals} 
\usepackage{graphics,epsfig}

\begin{document}

\title{\bf Mutual information harvesting for circularly accelerated detectors}
\author{
Mingkun Quan$^{1}$, Runhu Li$^{1}$, Zixu Zhao\footnote{Corresponding author. zhao$_{-}$zixu@yeah.net}
}
\affiliation{
School of Science, Xi'an University of Posts and Telecommunications, Xi'an 710121, China
}

\begin{abstract}

We investigate the mutual information harvesting of two circularly accelerated detectors that interact with the massless scalar fields near a reflecting boundary. We consider that the two detectors share a common rotational axis with the same acceleration and trajectory radius. As the interdetector separation increases, the mutual information may exhibit oscillatory behavior at large acceleration and small radius. For a fixed radius, a larger acceleration leads to a larger peak value of the mutual information. Near the boundary, the mutual information may oscillate and the maximum can be obtained. As the acceleration increases, the mutual information in a small interdetector separation first increases and then decreases. For an intermediate interdetector separation, the mutual information may oscillate with the increase of acceleration. For a not large interdetector separation, when we take large acceleration and small radius, as the energy gap increases, the mutual information first decreases, then oscillates, and finally goes to zero. The combination of large acceleration and small radius corresponds to the fast rotation, which significantly modifies the vacuum fluctuations of the field, leading to the oscillatory behavior. Furthermore, the oscillation intensifies near the boundary, which indicates that it is related to the coherent superposition of boundary reflections.

\end{abstract}

\maketitle

\section{Introduction}

In 1927, von Neumann gave the quantum analogue of the well known classical formula for the entropy \cite{Neumann1927}. In 1948, Shannon introduced the concept of information entropy \cite{Shannon1948}.  Subsequently, Fano introduced the term ``mutual information'' \cite{Kreer1957}. Entropy is one of the most important quantities in physics, Wehrl therefore discussed in detail general properties of entropy in 1978 \cite{Wehrl1978}. Cerf and Adami constructed a von Neumann mutual entropy based on a mutual amplitude operator \cite{Cerf1997}. Vedral et al. presented conditions every measure of entanglement has to satisfy, and constructed a whole class of ``good'' entanglement measures by generalizing von Neumann mutual information \cite{Vedral1997}.

It is of great interest to consider the properties of noninertial detectors. In 1976, Unruh examined the behavior of particle detectors under accelerated motion and applied these results to the black hole evaporation problem \cite{Unruh1976}. In 2003, Alsing considered quantum entanglement in noninertial frames, and showed that the fidelity of teleportation is reduced when the receiver is making observations in a uniformly accelerated frame \cite{Alsing2003}. In 2004, Peres and Terno discussed the intimate relationship between quantum mechanics, information theory, and relativity theory \cite{Peres2004}. In 2005, Fuentes-Schuller and Mann showed that a state which is maximally entangled in an inertial frame becomes less entangled if the observers are relatively accelerated \cite{Schuller2005}. In 2009, Landulfo and Matsas calculated the mutual information and concurrence between the two qubits and showed that the latter has a ``sudden death'' at a finite acceleration, whose value will depend on the time interval along which the detector is accelerated \cite{Landulfo2009}. Relativistic quantum information has become a new field of high research intensity, which aims to understand the relationship between special relativity, general relativity, and quantum information \cite{Mann2012}.

Because the vacuum can be a resource, Summers and Werner found the vacuum state in any Bose or Fermi free quantum field theory violates Bell's inequalities maximally \cite{Summers}. Valentini predicted a non-locally correlated joint spontaneous emission, which can be understood either in terms of non-local photon ``propagation'', or in terms of non-locally-correlated vacuum-field fluctuations \cite{Valentini1991}. Reznik explored the entanglement of the vacuum of a relativistic field by letting a pair of causally disconnected probes interact with the field \cite{Reznik2003}. Reznik, Retzker and Silman used a pair of initially nonentangled detectors locally interact with the vacuum for a finite duration, such that the two detectors remain causally disconnected, and then analyzed the resulting detector mixed state \cite{Reznik2005}. Entanglement harvesting has been extensively studied \cite{Martin2016,Kukita2017,Henderson2018,Henderson2019,Cong2019,Koga2019,Zhang2020,Liu2021,Liu2022,Hu2022,Suryaatmadja2022,Li2025,Brown2013,Pozas-Kerstjens2015}.

Mutual information harvesting also has been explored in recent years \cite{Brown2013,Pozas-Kerstjens2015,Simidzija2018,Gallock2021,Sahu2022,Bueley2022,Liu2023,Naeem2023,Bozanic2023,Wang2025,Huang2025}. In 2013, Brown studied the mutual information locally harvested from a field \cite{Brown2013}. In 2015, Pozas-Kerstjens and Mart\'{\i}n-Mart\'{\i}nez performed a detailed study of the phenomenon of mutual information harvesting from the vacuum state of a scalar field, using a pair of Unruh-DeWitt particle detectors \cite{Pozas-Kerstjens2015}. In 2018, Simidzija and Mart\'{\i}n-Mart\'{\i}nez studied the harvesting of mutual information by Unruh-DeWitt particle detectors from thermal and squeezed coherent field states \cite{Simidzija2018}. In 2021, Gallock-Yoshimura and Mann found that initially separable or weakly entangled detectors can extract mutual information from the vacuum \cite{Gallock2021}. In 2023, Liu, Zhang and Yu studied the phenomena of mutual information harvesting for two static detectors with a boundary \cite{Liu2023}. Naeem et al. investigated the mutual information harvesting protocol for two uniformly accelerated particle detectors \cite{Naeem2023}. In this paper, we study the mutual information harvesting for two circularly accelerated detectors near a boundary.

The organization of the work is as follows. In Sec. II, we review the UDW detector model and introduce the conventional measures for mutual information. In Sec. III, we study the phenomena of mutual information harvesting for a pair of coaxial accelerating detectors moving along a circular trajectory in the presence of a reflecting boundary. We conclude in the last section with our main results. We adopt natural units $\hbar = c = 1$ for simplicity.

\section{The basic formalism}\label{sec:basic formalism}

In this section, we consider a pair of detectors $A$ and $B$ interacting locally with a quantum scalar field $\phi[x_D(\tau)]$ {($D\in\{A,B\}$)}. The spacetime trajectory $x_D(\tau)$ of the detector is parameterized in terms of its proper time. The interacting Hamiltonian for such a detector locally coupling with a massless scalar field has the following form
\begin{equation}\label{Int1}
H_{D}(\tau)=\lambda \chi(\tau)\left[e^{i \Omega_{D}\tau} \sigma^{+}+e^{-i \Omega_{D}\tau} \sigma^{-}\right] \phi\left[x_{D}(\tau)\right],~~ D\in\{A,B\}\;,
\end{equation}
where $\lambda$ denotes the coupling strength and $\chi(\tau)=\exp[-{\tau^{2}}/(2\sigma^{2})]$ is the Gaussian switching function which controls the duration of interaction via parameter $\sigma$. The two-level atom with the ground state $\ket{0}_D$ and excited state $\ket{1}_D$ separated by an energy gap $\Omega_D$ can be modeled with the UDW detector. Particularly, $\sigma^{+}=|1\rangle_{D}\langle0|_{D}$ and $\sigma^{-}=|0\rangle_{D}\langle1|_{D}$ correspond to the raising and lowering operators, respectively.

We assume that the two detectors are initially prepared in their ground states and the field is in the Minkowski vacuum state $|0\rangle_{M}$. Therefore the initial joint state of the detectors and the field can be written as $|\Psi\rangle_{i} = |0\rangle_{A}|0\rangle_{B}|0\rangle_{M}$. Using the Hamiltonian~(\ref{Int1}), the time evolution of the quantum system can be written as
\begin{equation}\label{psi-f}
|\Psi\rangle_{f}:={\cal{T}} \exp\Big[-i\int{dt}\Big(\frac{d\tau_A}{dt}H_A(\tau_A)+\frac{d\tau_B}{dt}{H_B}(\tau_B)\Big)\Big]|\Psi\rangle_{i}\;,
\end{equation}
here ${\cal{T}}$ denotes the time-ordering operator, and $t$ is the coordinate time with respect to which the vacuum state of the field is defined. In the basis $\{|0\rangle_{A}|0\rangle_{B},|0\rangle_{A}|1\rangle_{B},|1\rangle_{A}|0\rangle_{B}, |1\rangle_{A}|1\rangle_{B}\}$, the final reduced density matrix of the two detectors can be obtained by tracing out the field degrees of freedom \cite{Pozas-Kerstjens2015}
\begin{align}\label{rhoAB}
\rho_{AB}:&=tr_{\phi}\big(U \left| {\Psi_{f}}\right\rangle \left\langle {\Psi_{f}}\right|  U^{\dag}\big)\nonumber\\
&=\begin{pmatrix}
1-P_A-P_B & 0 & 0 & X \\
0 & P_B & C & 0 \\
0 & C^* & P_A & 0 \\
X^* & 0 & 0 & 0 \\
\end{pmatrix}+{\mathcal{O}}(\lambda^4)\;.
\end{align}
\begin{equation}\label{PAPB}
P_D:=\lambda^{2}\iint d\tau d\tau' \chi(\tau) \chi(\tau') e^{-i \Omega_{D}(\tau-\tau')}
W\left(x_D(t), x_D(t')\right)\quad\quad D\in\{A, B\}\;,
\end{equation}
and the quantities $C$ and $X$ read
\begin{align}\label{ccdef}
C:=&\lambda^2 \iint dt  dt' \,  \frac{\partial\tau_B}{\partial{t}} \frac{\partial\tau_A}{\partial{t'}} \chi_B(\tau_B(t))  \chi_A(\tau_A(t')) e^{i \left[ \Omega_B\tau_B(t)-\Omega_A\tau_A(t'\right)]} W\!\left(x_A(t') , x_B(t)\right)\;,
\end{align}
\begin{align}\label{xxdef}
X:=-\lambda^2  \iint_{t>t'} dtdt'&\bigg[
\frac{\partial\tau_B}{\partial{t}} \frac{\partial\tau_A}{\partial{t'}} \chi_B(\tau_B(t)) \chi_A(\tau_A(t'))e^{-i\left[\Omega_B\tau_B(t)+\Omega_A\tau_A(t')\right]}W\!\left(x_A(t'), x_B(t)\right)  \notag\\
& +\frac{\partial\tau_A}{\partial{t}}\frac{\partial\tau_B}{\partial{t'}} \chi_A(\tau_A(t)) \chi_B(\tau_B(t'))e^{-i\left[\Omega_A\tau_A(t)+\Omega_B\tau_B(t')\right]}W\!\left(x_B(t'),x_A(t)\right)\bigg],
\end{align}
where $W(x,x'):=\langle0|_{M}\phi(x)\phi(x')|0\rangle_{M}$ denotes the Wightman function of the quantum field in the Minkowski vacuum state. The total amount of correlations can be measured by the mutual information, which is given by \cite{Pozas-Kerstjens2015}
\begin{equation}\label{IIdf0}
\mathcal{I}(\rho_{AB})= S(\rho_{A})+S(\rho_{B})-S(\rho_{AB})\;,
\end{equation}
here, $\rho_{A}=\operatorname{tr}_{B}(\rho_{AB})$ and $\rho_{B}=\operatorname{tr}_{A}(\rho_{AB})$ represent the reduced density matrices of detectors $A$ and $B$, respectively, obtained by tracing over the degrees of freedom of the other detector from the joint state $\rho_{AB}$. The function $S(\rho) = -\operatorname{tr}(\rho \ln \rho)$ denotes the von Neumann entropy. Utilizing the above definition~(\ref{IIdf0}), the mutual information for the final state of the detectors can be written as
\begin{align}\label{IIdf}
\mathcal{I}(\rho_{AB})=&\mathcal{L}_+\ln(\mathcal{L}_{+})+\mathcal{L}_-\ln(\mathcal{L}_{-})-P_{A}\ln(P_{A})-P_{B}\ln (P_{B})+\mathcal{O}(\lambda^4),
\end{align}
with
\begin{align}
\mathcal{L}_\pm&=\frac{1}{2}\Big(P_{A}+P_{B}\pm\sqrt{ (P_{A}-P_{B})^2 + 4 |C|^2 }\Big)\;.
\end{align}
The mutual information ${\cal{I}}(\rho_{AB})$ is determined by the transition probabilities $P_{A}$, $P_{B}$, and the correlation term $C$. From Eq.~(\ref{IIdf}), we can conclude that ${\cal{I}}(\rho_{AB}) = 0$ if $C = 0$. In particular, when one of the transition probabilities is zero, the mutual information must vanish due to the positivity condition of the density matrix (i.e., $P_{A}P_{B} \geq |C|^2$).

\section{Mutual information harvesting for Unruh-DeWitt detectors in circular motion with a reflecting boundary}\label{sec:Mutual information harvesting}

\begin{figure}[!htbp]
\centering
\includegraphics[width=0.6\textwidth]{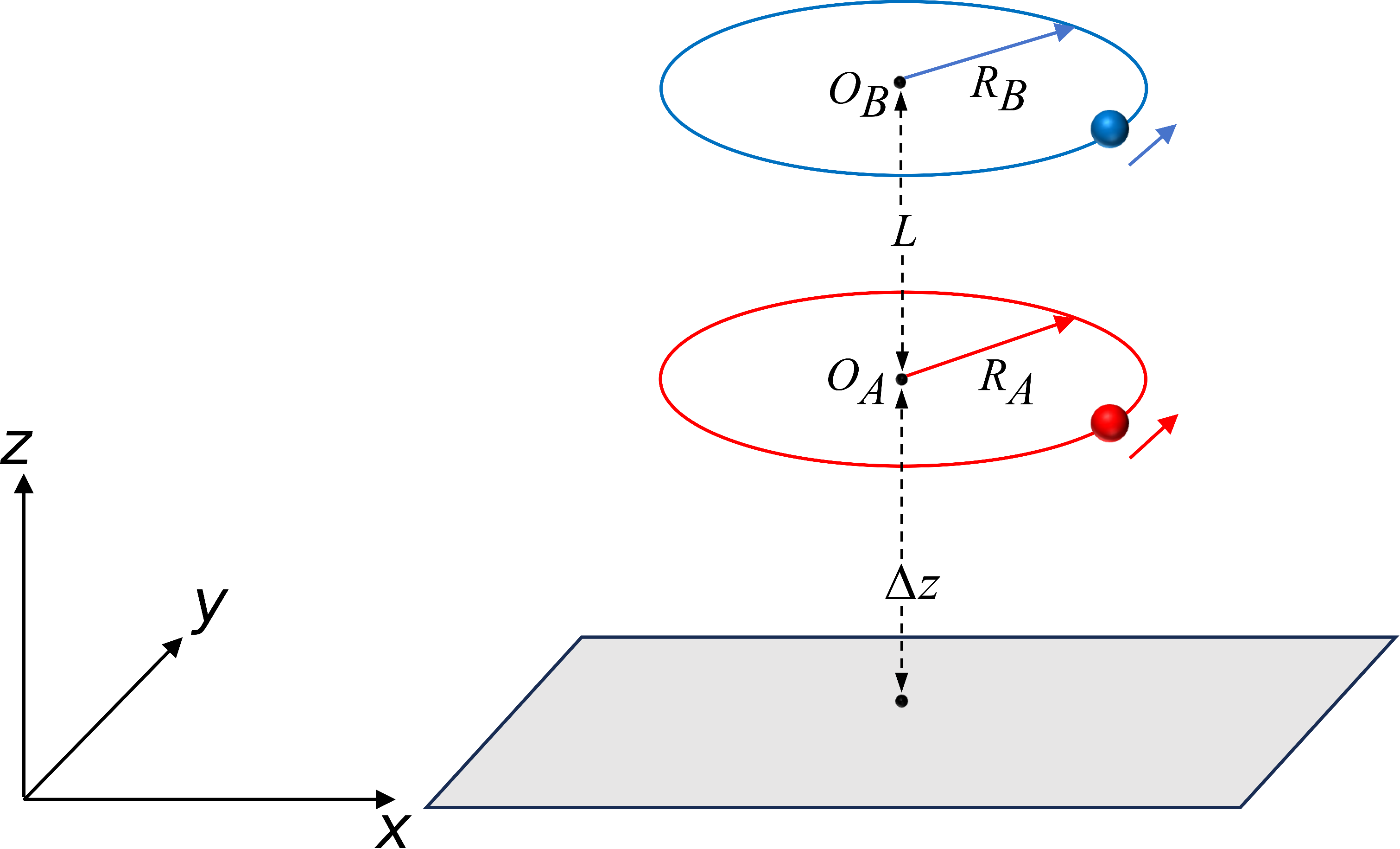}
\caption{The circular motion of two UDW detectors A and B is considered in flat spacetime. The boundary lies in the $xy$-plane and the two detectors  rotate around a common axis.}
\label{trajectory}
\end{figure}

In this section, we study the mutual information harvesting of two circularly accelerated detectors with a boundary. For simplicity, we focus on the spacetime trajectory of the detectors during coaxial rotation (see Fig.~\ref{trajectory}). We assume that the planar boundary is located at $z=0$, and the accelerated UDW detectors move along the circular trajectories parallel to the $xy$-plane with a  distance $\Delta z$ from the boundary.

We consider detectors $A$ and $B$ rotating around a common axis with angular velocities $\omega_A$ and $\omega_B$, and radii $R_A$ and $R_B$. Their spacetime trajectories are parameterized by the proper time $\tau_A$ and $\tau_B$
\begin{align}\label{traj2}
&x_A:=\{t=\tau_A\gamma_A\;~,~x=R_A\cos(\omega_A\tau_A\gamma_A)\;~,~
y=R_A\sin(\omega_A\tau_A\gamma_A)\;~,~z=\Delta z\}\;,\nonumber\\
&x_B:=\{t=\tau_B\gamma_B\;~,~x=R_B\cos(\omega_B\tau_B\gamma_B)\;~,~
y=R_B\sin(\omega_B\tau_B\gamma_B)\;~,~z=\Delta z+L\}\;.
\end{align}
Here, $\gamma_A$ and $\gamma_B$ are corresponding Lorentz factors for detectors $A$ and $B$, and $L$ denotes the distance between the two detectors.

The Wightman function for vacuum massless scalar fields with a reflecting boundary is given by \cite{Birrell1982}
\begin{equation}\label{wightman1}
\begin{aligned}
W(x,x')=&-\frac{1}{4\pi^2}\big[\frac{1}{(t-t'-i\epsilon)^2-(x-x')^2-(y-y')^2-(z-z')^2}\\
&-\frac{1}{(t-t'-i\epsilon)^2-(x-x')^2-(y-y')^2-(z+z')^2}\big]\;.
\end{aligned}
\end{equation}

\begin{figure*}[!htbp]
\centering
\subfigure[$a\sigma=0.10, R/\sigma=0.02$]{\label{IvsLz01_a}
\includegraphics[width=0.3\textwidth]{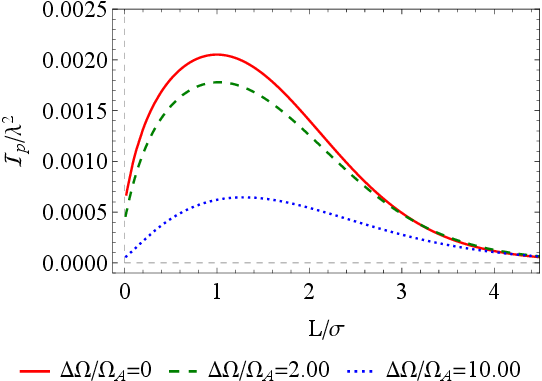}}\hspace{0.005\textwidth}
\subfigure[$a\sigma=1.00, R/\sigma=0.02$]{\label{IvsLz01_b}
\includegraphics[width=0.3\textwidth]{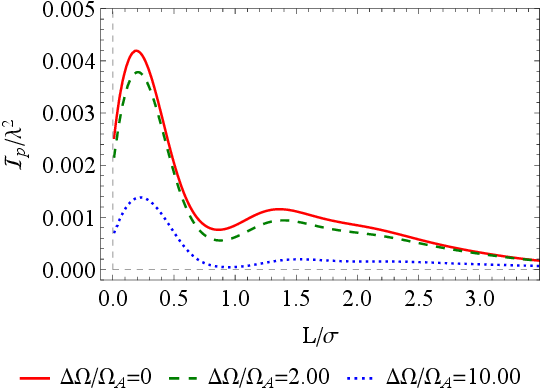}}\hspace{0.005\textwidth}
\subfigure[$a\sigma=5.00, R/\sigma=0.02$]{\label{IvsLz01_c}
\includegraphics[width=0.3\textwidth]{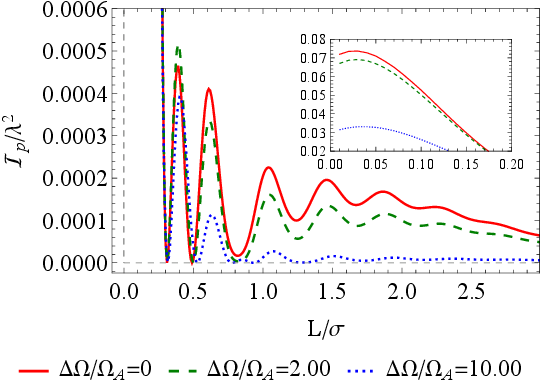}}
\subfigure[$a\sigma=0.10, R/\sigma=10.00$]{\label{IvsLz01_d}
\includegraphics[width=0.3\textwidth]{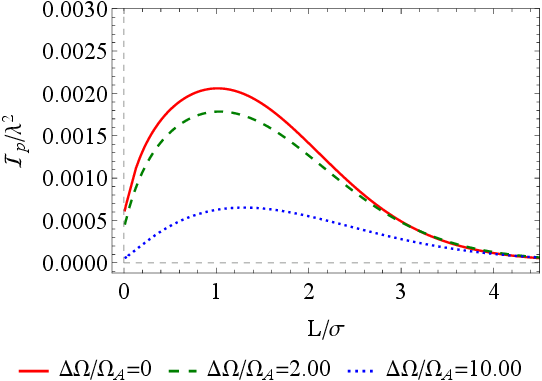}}\hspace{0.005\textwidth}
\subfigure[$a\sigma=1.00, R/\sigma=10.00$]{\label{IvsLz01_e}
\includegraphics[width=0.3\textwidth]{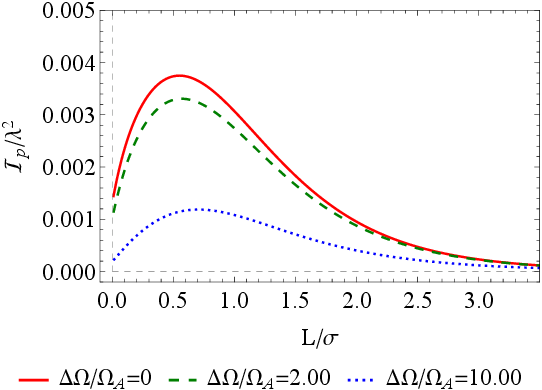}}\hspace{0.005\textwidth}
\subfigure[$a\sigma=5.00, R/\sigma=10.00$]{\label{IvsLz01_f}
\includegraphics[width=0.3\textwidth]{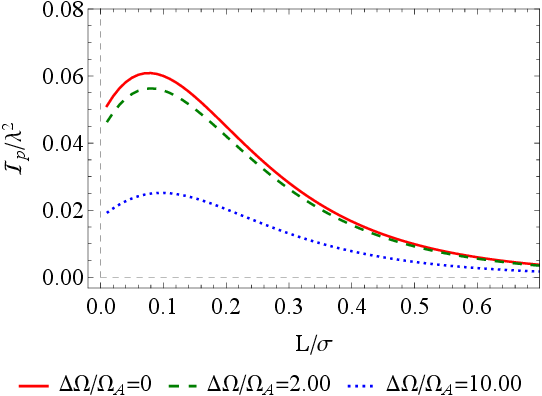}}
\caption{The mutual information $I_p/\lambda^2$ is plotted as a function of interdetector separation $L/\sigma$ for different $a\sigma$ and $R/\sigma$. We have set $\Omega_A\sigma=0.10$, $\Delta\Omega/\Omega_A=\left\{0, 2.00, 10.00 \right\}$ and $\Delta z/\sigma=0.10$.}
\label{IvsLz01}
\end{figure*}

Substituting the trajectories~(\ref{traj2}) and Eq.~(\ref{wightman1}) into Eq.~(\ref{PAPB}), the transition probabilities can be obtained \cite{Li2025}
\begin{align}\label{PD}
\begin{aligned}
P_D=&K_D\int_0^{\infty }dx\frac{e^{-\alpha x^2}\cos (\beta x)(x^2-\sin^2 x)}{x^2 (x^2-v_D ^2 \sin^2 x)}\, +\frac{\lambda ^2 |\omega _{D}|\sigma}{4\pi ^{3/2} \gamma}{\rm{PV}} \int_0^{\infty }dx \frac{e^{-\alpha  x^2} \cos (\beta x)}{x^2-v_{D}^2\sin ^2(x)-\omega _{D}^2\Delta z^2 } \, \\
&+\frac{\lambda^2}{4\pi}\Big[e^{-\Omega^2\sigma^2}-\sqrt{\pi}\Omega\sigma {\rm{Erfc}}\big(\Omega\sigma\big)\Big]+\frac{\lambda ^2|\omega _{D}| \sigma}{4 \sqrt{\pi} \gamma}\frac{e^{-\alpha  S^2} \sin (\beta  S)}{2S-v_{D}^2 \sin (2S)},
\end{aligned}
\end{align}
where
\begin{equation}
\alpha=\frac{1}{\sigma^2\omega_D^2\gamma_D^2}=\frac{R_D}{a_D\sigma^2}\;,~~
\beta=\frac{2\Omega}{\gamma_D|\omega_D|}\;,~~K_D=\frac{\lambda^2v_D^2\gamma_D|\omega_D|\sigma}{4\pi^{3/2}}=\frac{\lambda^2v_Da_D\sigma}{4\pi^{3/2}\gamma_D}\;,
\end{equation}
and ${\rm{Erfc}}(x)=1-{\rm{Erf}}(x)$ is the complementary error function with ${\rm{Erf}}(x):=\int_{0}^{x} 2e^{-t^2}dt/\sqrt{\pi}$. $S$ is the solution of the equation $x^2-v_{D}^2\sin^2x-\omega_D^2\Delta z^2=0$.

Substituting the trajectories Eq.~(\ref{traj2}) and the Wightman function Eq.~(\ref{wightman1}) into Eq.~(\ref{ccdef}), we can obtain the correlation term
\begin{align}\label{C1}
C=\frac{\lambda^2\sigma^2}{4\pi^2\gamma_A\gamma_B}\int_{-\infty}^{\infty} &d\tilde{u} \int_{-\infty}^{\infty} d\tilde{s}\exp\Big[\frac{-\gamma_A^2\tilde{u}^2-\gamma_B^2(\tilde{u}-\tilde{s})^2}{2\gamma_A^2\gamma_B^2}\Big]\exp\Big[\frac{i\tilde{u}\sigma\Omega_B}{\gamma_B}-\frac{i(\tilde{u}-\tilde{s})\sigma\Omega_A}{\gamma_A}\Big]\nonumber\\
&\times(\frac{1}{f_{AB}(\tilde{u},\tilde{s})}-\frac{1}{f_{AB}(\tilde{u},\tilde{s})+4L\Delta{z}+4\Delta{z}^2}),
\end{align}
here the auxiliary functions read
\begin{equation}
f_{AB}(\tilde{u},\tilde{s})= L^2+R_A^2+R_B^2-2R_AR_B\cos\big({\tilde{u}\omega_A\sigma}-{\tilde{u}\omega_B\sigma-\tilde{s}\omega_A\sigma}\big)-\sigma^2(\tilde{s}+i\epsilon)^2\;.
\end{equation}

For simplicity, we assume that the two detectors are rotating synchronously around the $z$-axis with the same acceleration and orbital radius, i.e., $a_A=a_B=a$, $R_A=R_B=R$ (or equivalently, $v_A=v_B,\; \omega_A=\omega_B$). In this case, it is straightforward to determine that $ \gamma_A=\gamma_B=\gamma $. The correlation Eq.~(\ref{C1}) can then be written as
\begin{align}\label{C2}
C = \frac{\lambda^2 \sigma^2}{4 \pi^{3/2} \gamma} e^{-\frac{1}{4} \Delta\Omega^2 \sigma^2} \int_{-\infty}^{\infty} d\tilde{s}\ e^{-\frac{\tilde{s}^2}{4 \gamma^2}} e^{\frac{i \tilde{s} \sigma (\Delta \Omega + 2 \Omega_A)}{2 \gamma}} ( \frac{1}{f_{AB}} - \frac{1}{f_{AB} +4L\Delta{z}+4\Delta{z}^2} ),
\end{align}
where $\Delta\Omega=\Omega_B-\Omega_A$ and
\begin{equation}
f_{AB}=L^2+4R^2\sin^2(\tilde{s}\omega\sigma/2)-\sigma^2(\tilde{s}+i\epsilon)^2\;.
\end{equation}

Eq.~(\ref{C2}) can be expressed as
\begin{align}\label{c0}
C = C_1-C_2,
\end{align}
with
\begin{align}\label{c1}
C_1= \frac{\lambda^2 \sigma^2}{4 \pi^{3/2} \gamma} e^{-\frac{1}{4} \Delta\Omega^2 \sigma^2} \int_{-\infty}^{\infty} d\tilde{s}\ e^{-\frac{\tilde{s}^2}{4 \gamma^2}} e^{\frac{i \tilde{s} \sigma (\Delta \Omega + 2 \Omega_A)}{2 \gamma}} \frac{1}{f_{AB}},
\end{align}
and
\begin{align}\label{c2}	C_2= \frac{\lambda^2 \sigma^2}{4 \pi^{3/2} \gamma} e^{-\frac{1}{4} \Delta\Omega^2 \sigma^2} \int_{-\infty}^{\infty} d\tilde{s}\ e^{-\frac{\tilde{s}^2}{4 \gamma^2}} e^{\frac{i \tilde{s} \sigma (\Delta \Omega + 2 \Omega_A)}{2 \gamma}} \frac{1}{f_{AB} +4L\Delta{z}+4\Delta{z}^2}.
\end{align}
$C_1$ corresponds to the correlation term of the circularly accelerated detectors in free Minkowski spacetime, and $C_2$ arises from the presence of the boundary. Because analytical results are quite difficult to obtain, we employ numerical calculations \cite{Naeem2023}.

The existence of a certain critical value will lead to a transition in the behavior of mutual information. To obtain more mutual information, we take a smaller energy gap. Taking $\Omega_A\sigma=0.10$, there exists a critical value $\Delta z_c/\sigma \approx 2.739$. When $\Delta z/\sigma < \Delta z_c/\sigma$, we display the mutual information as a function of interdetector separation $L/\sigma$ for a small $\Delta z/\sigma =0.10$ in Fig.~\ref{IvsLz01}. We first consider a small $R/\sigma=0.02$. When we take a small $a\sigma=0.10$, as $L/\sigma$ increases, $I_p/\lambda^2$ initially increases, then decreases, and finally goes to zero, which is different from the case of static detectors with a boundary \cite{Liu2023}. A single peak appears at a specific interdetector separation for each energy gap. For an intermediate $a\sigma=1.00$, there exists two peaks for each curve. For a large $a\sigma=5.00$, there exist oscillatory behavior. The oscillatory behavior appears for the combination of large $a\sigma$ and small $R/\sigma$, which stands for fast rotation and indicates that the oscillation is related to the periodic nature of circular motion. When a large $R/\sigma=10.00$ is considered, for any $a\sigma$, $I_p/\lambda^2$ first increases, then decreases. Moreover, for fixed $R/\sigma$, a larger acceleration leads to a higher peak value of the mutual information.

\begin{figure*}[!ht]
\centering
\subfigure[$a\sigma=0.10, R/\sigma=0.02$]{\label{IvsLz5_a}
\includegraphics[width=0.4\textwidth]{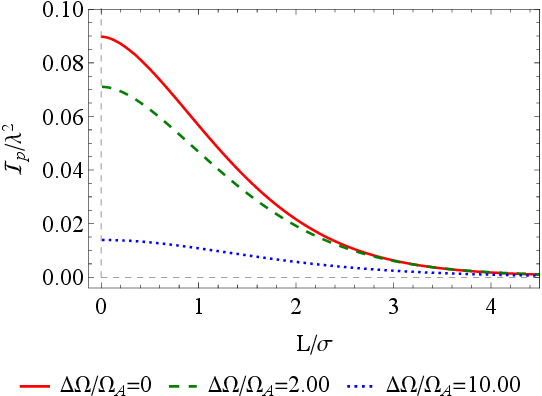}}\hspace{0.02\textwidth}
\subfigure[$a\sigma=5.00, R/\sigma=0.02$]{\label{IvsLz5_b}
\includegraphics[width=0.4\textwidth]{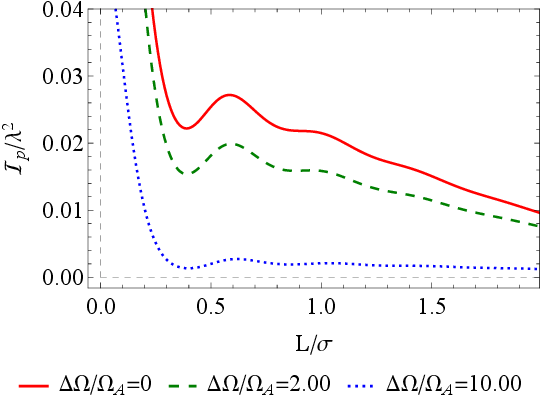}}
\subfigure[$a\sigma=0.10, R/\sigma=10.00$]{\label{IvsLz5_c}
\includegraphics[width=0.4\textwidth]{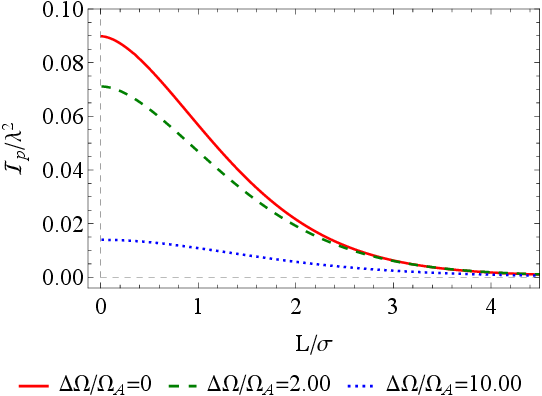}}\hspace{0.02\textwidth}
\subfigure[$a\sigma=5.00, R/\sigma=10.00$]{\label{IvsLz5_d}
\includegraphics[width=0.4\textwidth]{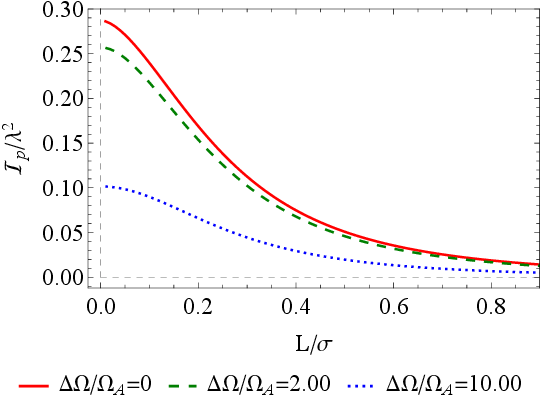}}
\caption{The mutual information $I_p/\lambda^2$ as a function of the interdetector separation $L/\sigma$ is plotted for different $a\sigma$ and $R/\sigma$. We have set $\Omega_A\sigma=0.10$, $\Delta\Omega/\Omega_A=\left\{0, 2.00, 10.00 \right\}$ and $\Delta z/\sigma=5.00$.}
\label{IvsLz5}
\end{figure*}

\begin{figure}[!htbp]
\centering
\includegraphics[width=0.4\textwidth]{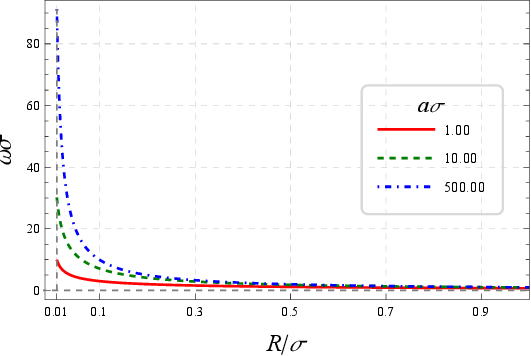}
\caption{Angular velocity $\omega\sigma$ as a function of $R/\sigma$ is plotted. We have set $a\sigma=\left\{1.00, 10.00, 500.00 \right\}$.}
\label{omegavsa}
\end{figure}

\begin{figure}[!htbp]
\centering
\begin{tabular}{c@{\hspace{1cm}}c}
\includegraphics[width=0.4\textwidth]{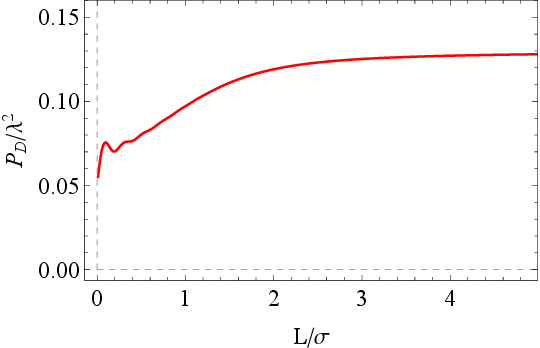} &
\includegraphics[width=0.4\textwidth]{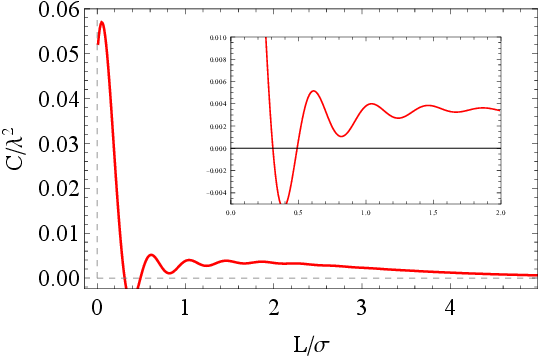}\\
\end{tabular}
\caption{$P_{D}/\lambda^2$ and $C/\lambda^2$ are plotted as functions of $L/\sigma$. We have set  $\Omega_B\sigma=\Omega_A\sigma=0.10$, $a\sigma=5.00$, $R/\sigma=0.02$, and $\Delta z/\sigma=0.10$.}
\label{CPvsL}
\end{figure}

\begin{figure}[!htbp]
\centering
\begin{tabular}{c@{\hspace{1cm}}c}
\includegraphics[width=0.4\textwidth]{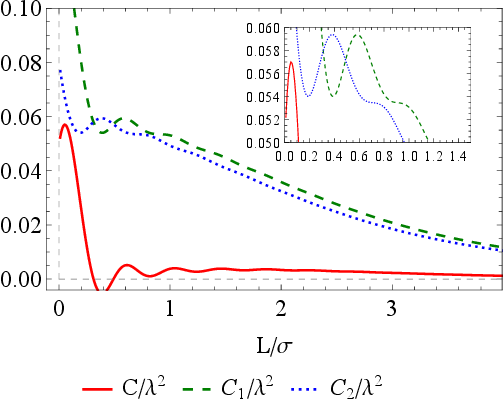} &
\includegraphics[width=0.4\textwidth]{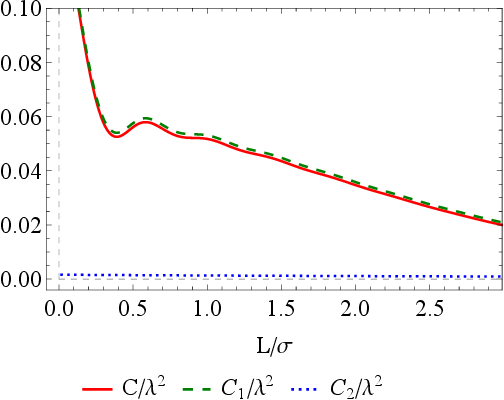} \\
\small (a) $\Delta z/\sigma=0.10$  & \small (b) $\Delta z/\sigma=5.00$  \\
\end{tabular}
\caption{$C/\lambda^2$, $C_1/\lambda^2$ and $C_2/\lambda^2$ are plotted as functions of $L/\sigma$ for different $\Delta z/\sigma$. We have set $\Omega_B\sigma=\Omega_A\sigma=0.10$, $a\sigma=5.00$, $R/\sigma=0.02$.}
\label{C12vsL}
\end{figure}

\begin{figure}[!htbp]
\centering
\begin{tabular}{c@{\hspace{1cm}}c}
\includegraphics[width=0.4\textwidth]{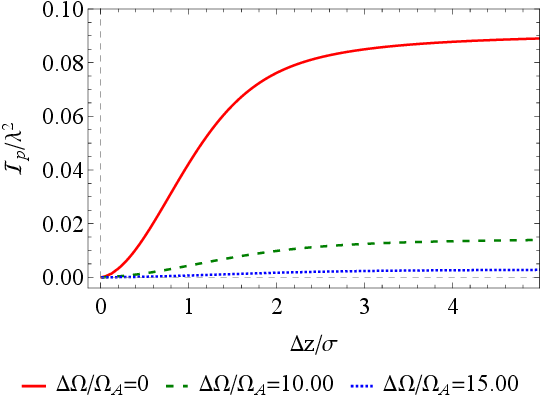} &
\includegraphics[width=0.4\textwidth]{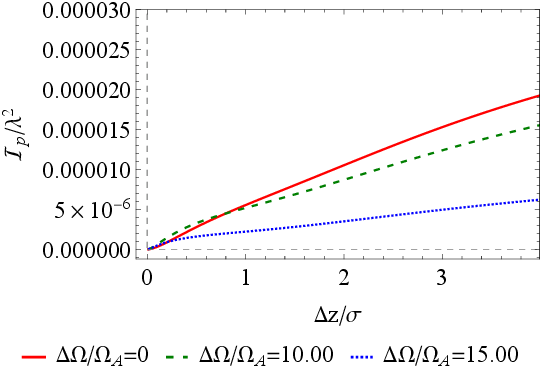}\\
\small (a) $a\sigma=0.10$, $L/\sigma=0.10$ & \small (b) $a\sigma=0.10$, $L/\sigma=10.00$  \\	
\includegraphics[width=0.4\textwidth]{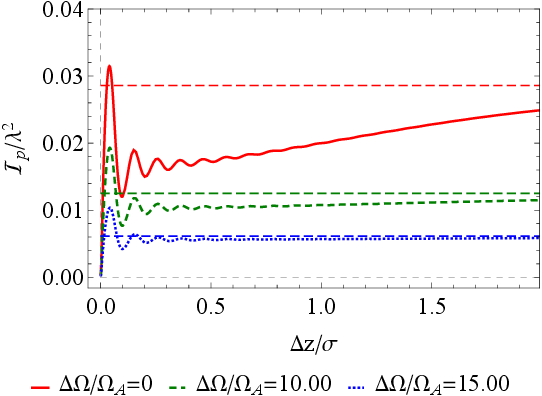} &
\includegraphics[width=0.4\textwidth]{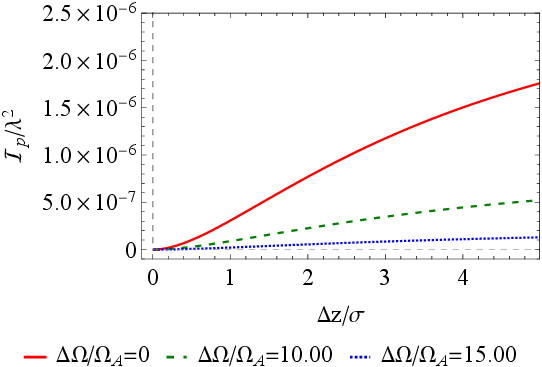}\\
\small (c) $a\sigma=30.00$, $L/\sigma=0.10$ & \small (d) $a\sigma=30.00$, $L/\sigma=10.00$  \\
\end{tabular}
\caption{The mutual information as a function of $\Delta z/\sigma $ is plotted for different $a\sigma$ and $L/\sigma$. We have set $\Omega_A\sigma=0.10$, $\Delta\Omega/\Omega_A=\left\{0, 10.00, 15.00 \right\}$ and $R/\sigma=0.02$. The corresponding results without boundary are shown as dashed straight lines.}
\label{Ivszr002}
\end{figure}

\begin{figure}[!htbp]
\centering
\begin{tabular}{c@{\hspace{1cm}}c}
\includegraphics[width=0.4\textwidth]{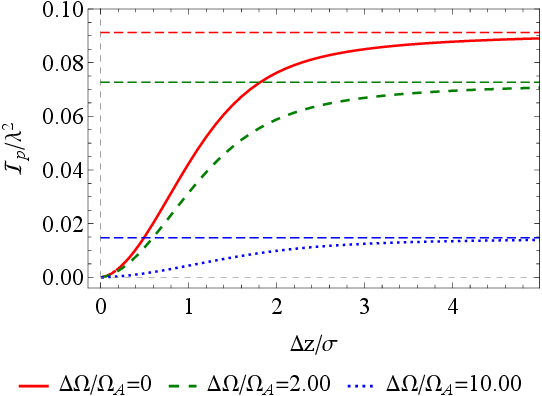} &
\includegraphics[width=0.4\textwidth]{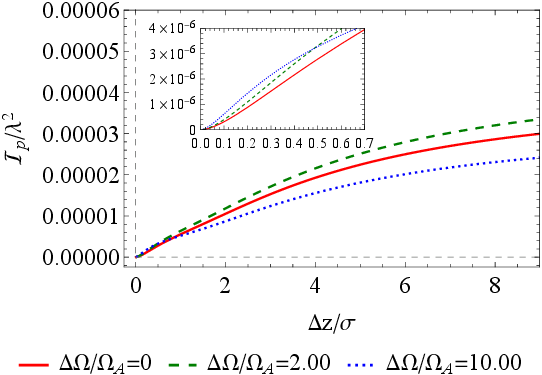}\\
\small (a) $a\sigma=0.10$, $L/\sigma=0.10$ & \small (b) $a\sigma=0.10$, $L/\sigma=10.00$  \\	
\includegraphics[width=0.4\textwidth]{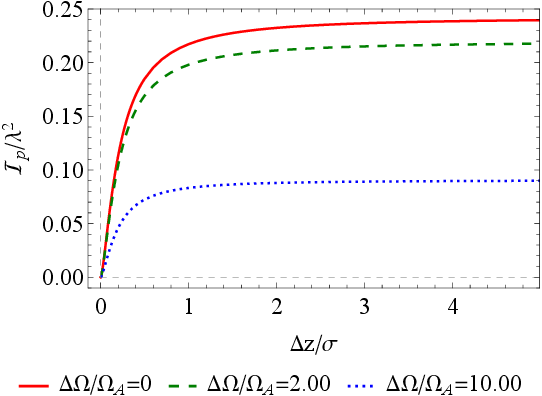} &
\includegraphics[width=0.4\textwidth]{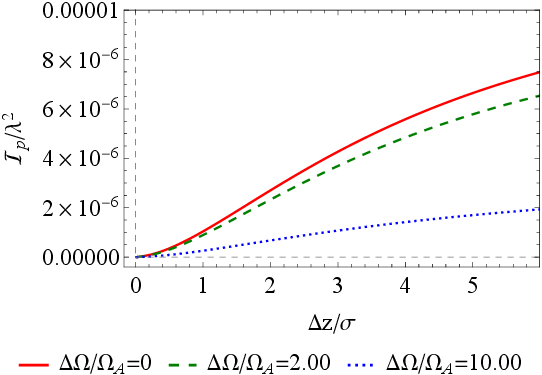}\\
\small (c) $a\sigma=5.00$, $L/\sigma=0.10$ & \small (d) $a\sigma=5.00$, $L/\sigma=10.00$  \\
\end{tabular}
\caption{The mutual information $I_p/\lambda^2$ is plotted as a function of $\Delta z/\sigma$ for different $a\sigma$ and $L/\sigma$. We have set $\Omega_A\sigma=0.10$, $\Delta\Omega/\Omega_A=\left\{0, 2.00, 10.00 \right\}$ and $R/\sigma=10.00$. The corresponding results in flat spacetime without any boundaries are shown as dashed straight lines.}
\label{Ivszr10}
\end{figure}

In Fig.~\ref{IvsLz5}, we present the mutual information as a function of interdetector separation $L/\sigma$ for a large $\Delta z/\sigma = 5.00 > \Delta z_c/\sigma$. The behavior is similar to the case without any boundary. For small $a\sigma=0.10$ or large $R/\sigma=10.00$, the mutual information $I_p/\lambda^2$ decreases monotonically as $L/\sigma$ increases, which is similar to the case of static detectors in the presence of a boundary \cite{Liu2023}. When we take a large $a\sigma=5.00$ and a small $R/\sigma=0.02$, $I_p/\lambda^2$ initially decreases, then exhibits oscillatory behavior, and eventually goes to zero. The oscillatory behavior is different from the linear acceleration case \cite{Naeem2023}.  The oscillation in Fig.~\ref{IvsLz01_c} is more intense compared to that in Fig.~\ref{IvsLz5_b}. This behavior indicates that the presence of the boundary enhances the oscillation intensity, which is related to the coherent superposition of boundary reflections. When $a\sigma>a_{c1}/\sigma$ ($a_{c1}/\sigma \approx 1.435$), the mutual information exhibits oscillatory behavior in Fig.~\ref{IvsLz01_c}. When $a\sigma>a_{c2}/\sigma$ ($a_{c2}/\sigma \approx 3.712$), the mutual information exhibits oscillatory behavior in Fig.~\ref{IvsLz5_b}. Therefore, we take $a\sigma=5.00$ in Fig.~\ref{IvsLz01_c} and Fig.~\ref{IvsLz5_b}.

From $\gamma = 1/\sqrt{1-R^2\omega^2}$ and $a=\gamma^2 \omega^2 R$, we have
\begin{equation}\label{omegavsar}
\begin{aligned}
\omega= \sqrt{\frac{a}{R(1+aR)}}\;.
\end{aligned}
\end{equation}
In Fig.~\ref{omegavsa}, we depict the angular velocity $\omega \sigma$ as a function of $R/\sigma $. We observe that a large $\omega \sigma$ corresponds to a combination of large acceleration and small radius. The denominator of the Wightman function contains the periodic factor $4R^2\sin^2(\tilde{s}\omega\sigma/2)$. Fast rotation corresponds to a large $\omega \sigma$ (i.e., large acceleration and small radius), which may lead to strong oscillation.
In Fig.~\ref{CPvsL}, we plot $P_{D}/\lambda^2$ and $C/\lambda^2$ as functions of $L/\sigma$. The oscillation is caused by a combined effect of $C$ and $P_D$.

In Fig.~\ref{C12vsL}, we plot $C/\lambda^2$, $C_1/\lambda^2$ and $C_2/\lambda^2$ with respect to $L/\sigma$. For a small $\Delta z/\sigma= 0.10$, we can observe strong oscillation due to the superposition of $C_1$ and $C_2$. For a large $\Delta z/\sigma= 5.00$, $C_2$ is close to zero, and the oscillation is weaker than that in the case of small $\Delta z/\sigma= 0.10$.

In Fig.~\ref{Ivszr002}, we describe the mutual information $I_p/\lambda^2$ as a function of $\Delta z/\sigma$ for small $R/\sigma = 0.02$. When we take a small $a\sigma=0.10$, the mutual information increases and finally tends to a stable value. This stable value corresponds to the unbounded case. For a large $L/\sigma=10.00$, there are intersections for different curves. When a large $a\sigma=30.00$ is considered, for a small $L/\sigma=0.10$, the mutual information initially increases, then oscillates, and eventually approaches to a stable value. For some parameters, the maximum value of the mutual information is higher than the stable value (i.e., unbounded case). This behavior indicates that large $a\sigma$ and small $R/\sigma$ encounter stronger vacuum fluctuations near the boundary. Therefore, we can harvest more mutual information by selecting appropriate parameters. The oscillation is also caused by a combined effect of $C$ and $P_D$. For a large $L/\sigma=10.00$, $I_p/\lambda^2$ increases and goes to a stable value.

We depict the mutual information $I_p/\lambda^2$ as a function of $\Delta z/\sigma $ for large $R/\sigma = 10.00$ in Fig.~\ref{Ivszr10}. The mutual information increases and eventually goes to a stable value. For a not large $L/\sigma=0.10$ or a large $a\sigma=5.00$, a larger energy gap difference corresponds to a smaller mutual information. For large $L/\sigma=10.00$ and small $a\sigma=0.10$, there are intersections for different curves.

\begin{figure}[!htbp]
\centering
\begin{tabular}{ccc}
\includegraphics[width=0.3\textwidth]{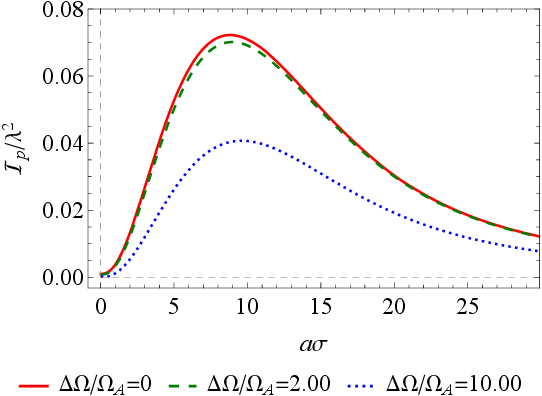} &
\includegraphics[width=0.3\textwidth]{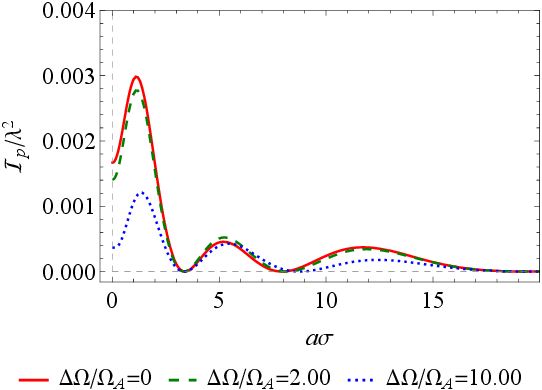} &
\includegraphics[width=0.3\textwidth]{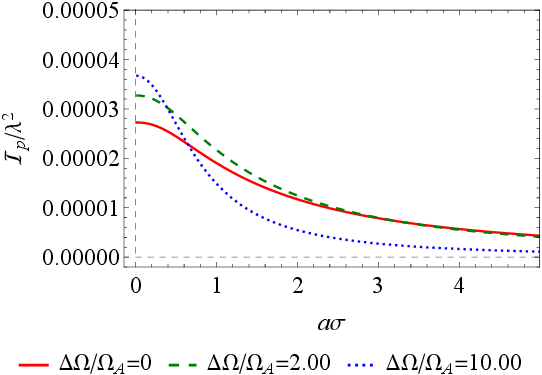} \\
\small (a) $L/\sigma=0.10$, $\Delta z/\sigma=0.10$  & \small (b) $L/\sigma=0.40$, $\Delta z/\sigma=0.10$  & \small (c) $L/\sigma=5.00$, $\Delta z/\sigma=0.10$ \\					
\includegraphics[width=0.3\textwidth]{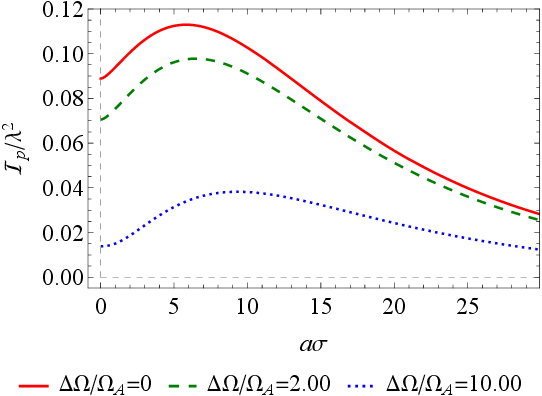} &
\includegraphics[width=0.3\textwidth]{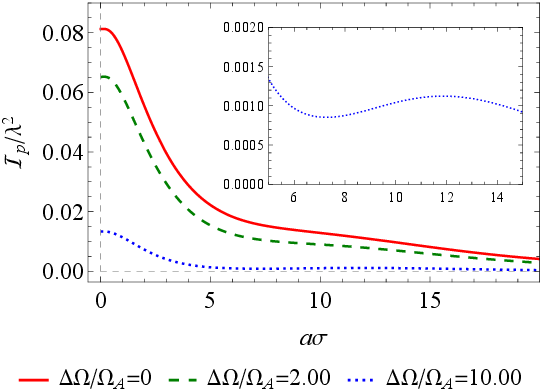} &
\includegraphics[width=0.3\textwidth]{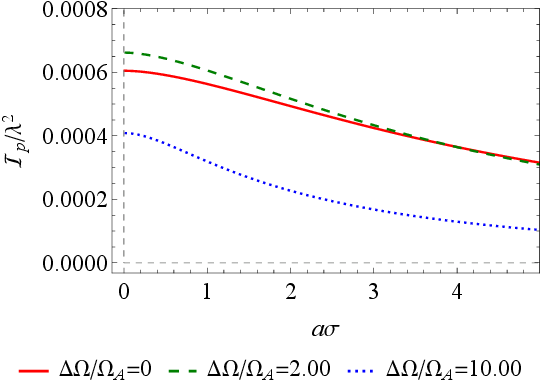} \\
\small (d) $L/\sigma=0.10$, $\Delta z/\sigma=5.00$  & \small (e) $L/\sigma=0.40$, $\Delta z/\sigma=5.00$  & \small (f) $L/\sigma=5.00$, $\Delta z/\sigma=5.00$ \\
\end{tabular}
\caption{The mutual information $I_p/\lambda^2$ is plotted as a function of $a\sigma$ for different $L/\sigma$ and $\Delta z/\sigma$. We have set $\Omega_A\sigma=0.10$, $\Delta\Omega/\Omega_A=\left\{0, 2.00, 10.00 \right\}$ and $R/\sigma=0.02$.}
\label{Ivsar002OA01}
\end{figure}

\begin{figure}[!htbp]
\centering
\begin{tabular}{ccc}
\includegraphics[width=0.3\textwidth]{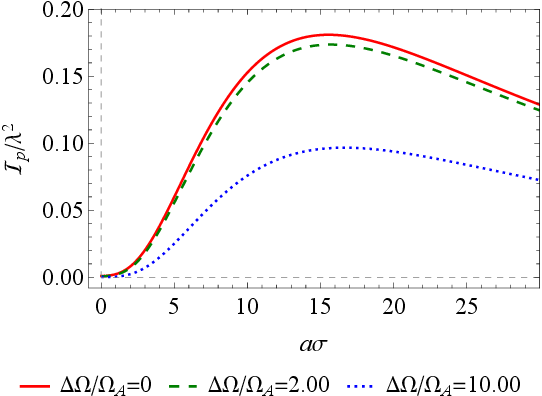} &
\includegraphics[width=0.3\textwidth]{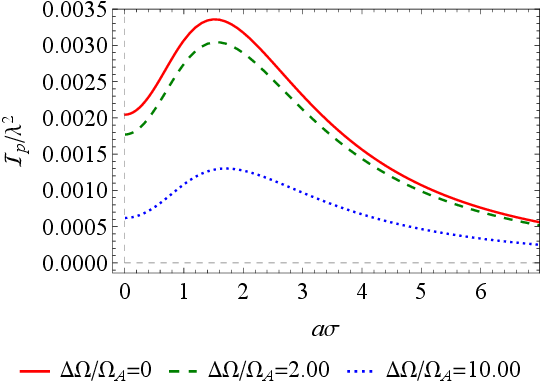} &
\includegraphics[width=0.3\textwidth]{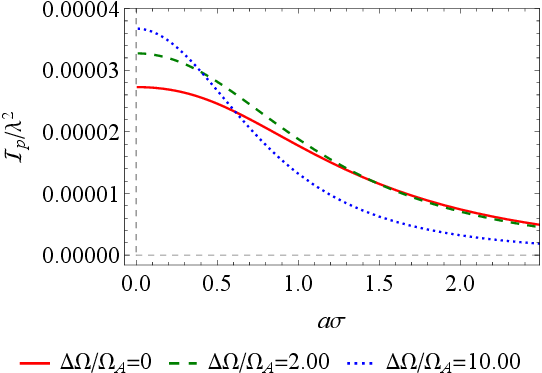} \\
\small (a) $L/\sigma=0.10$, $\Delta z/\sigma=0.10$  & \small (b) $L/\sigma=1.00$, $\Delta z/\sigma=0.10$  & \small (c) $L/\sigma=5.00$, $\Delta z/\sigma=0.10$ \\
\includegraphics[width=0.3\textwidth]{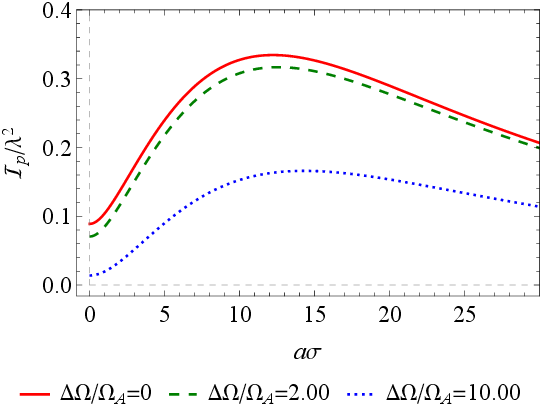} &
\includegraphics[width=0.3\textwidth]{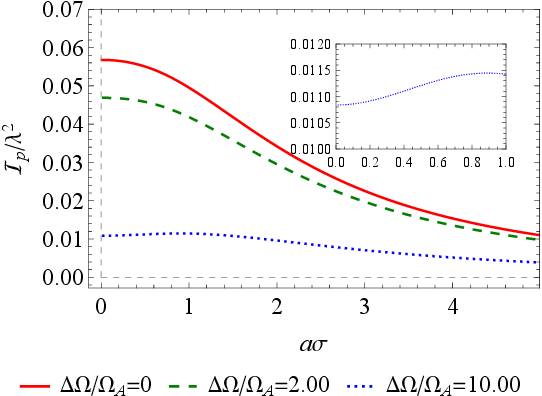} &
\includegraphics[width=0.3\textwidth]{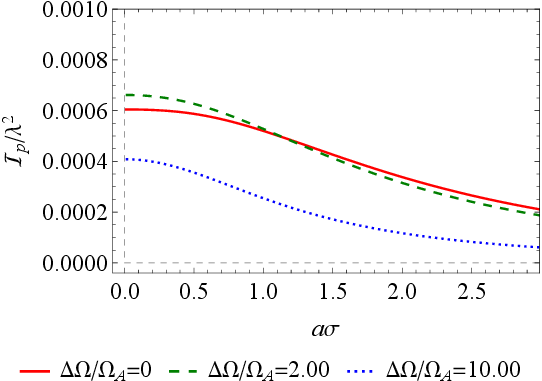} \\
\small (d) $L/\sigma=0.10$, $\Delta z/\sigma=5.00$  & \small (e) $L/\sigma=1.00$, $\Delta z/\sigma=5.00$  & \small (f) $L/\sigma=5.00$, $\Delta z/\sigma=5.00$ \\
\end{tabular}
\caption{The mutual information $I_p/\lambda^2$ is plotted as a function of $a\sigma$ for different $L/\sigma$ and $\Delta z/\sigma$. We have set $\Omega_A\sigma=0.10$, $\Delta\Omega/\Omega_A=\left\{0, 2.00, 10.00 \right\}$ and $R/\sigma=10.00$.}
\label{Ivsar10}
\end{figure}

In Fig.~\ref{Ivsar002OA01}, we plot the mutual information $I_p/\lambda^2$ as a function of $a\sigma$ for $R/\sigma = 0.02$. For a small $L/\sigma=0.10$, as $a\sigma$ increases, the mutual information $I_p/\lambda^2$ initially increases, and then decreases. We consider an intermediate $L/\sigma=0.40$, for a small $\Delta z/\sigma=0.10$, $I_p/\lambda^2$ first increases, then oscillates, and eventually approaches to zero. It should be noted that the oscillation comes from $C$, since the transition probability increases monotonically with the acceleration \cite{Zhang2020,Li2025}. For a large $\Delta z/\sigma$=5.00, the mutual information decreases non-monotonically for a large energy gap difference. When we take a large $L/\sigma=5.00$, there may exist intersections for different curves.

In Fig.~\ref{Ivsar10}, we plot the mutual information $I_p/\lambda^2$ as a function of $a\sigma$ for $R/\sigma = 10.00$. For a small $L/\sigma=0.10$, as $a\sigma$ increases, the mutual information $I_p/\lambda^2$ initially increases, and then decreases. When we consider an intermediate $L/\sigma=1.00$, for a small $\Delta z/\sigma=0.10$, $I_p/\lambda^2$ first increases, and then decreases. However, for a large $\Delta z/\sigma$=5.00, the mutual information decreases monotonically for small energy gap difference. For a larger energy gap difference, $I_p/\lambda^2$ increases initially, and then decreases. When a large $L/\sigma=5.00$ is considered, there may exist intersections for different curves.

\begin{figure}[!htbp]
\centering
\begin{tabular}{c@{\hspace{1cm}}c}
\includegraphics[width=0.4\textwidth]{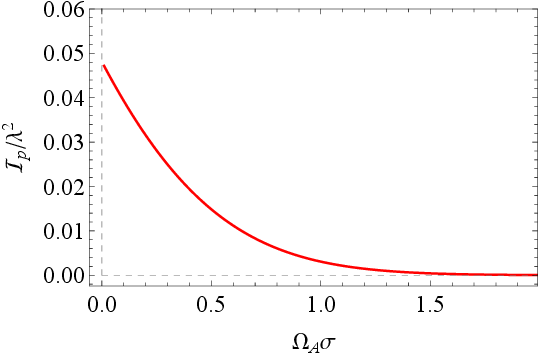} &
\includegraphics[width=0.4\textwidth]{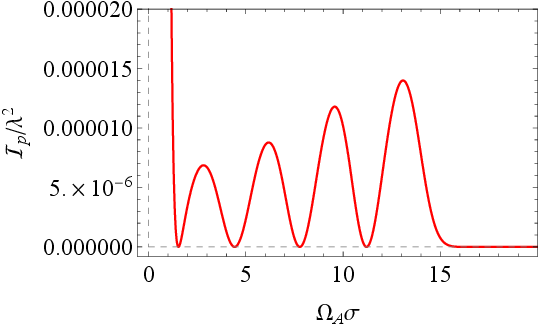} \\
\small (a) $a\sigma=0.10$, $R/\sigma=0.02$  &
\small (b) $a\sigma=5.00$, $R/\sigma=0.02$  \\		
\includegraphics[width=0.4\textwidth]{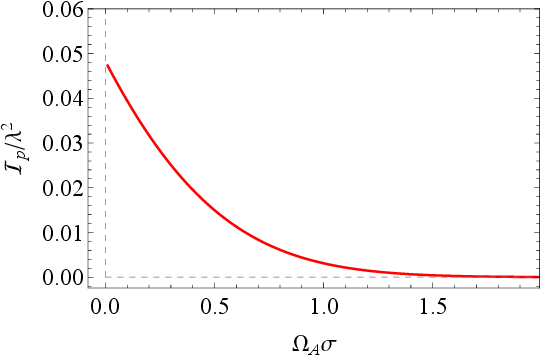} &
\includegraphics[width=0.4\textwidth]{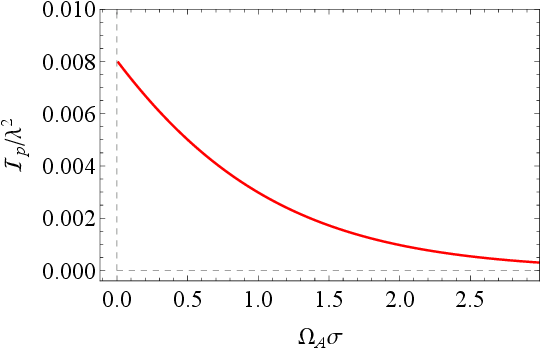} \\
\small (c) $a\sigma=0.10$, $R/\sigma=10.00$  &
\small (d) $a\sigma=5.00$, $R/\sigma=10.00$ \\
\end{tabular}
\caption{The mutual information $I_p/\lambda^2$ as a function of $\Omega_A\sigma $ is plotted for different $a\sigma$ and $R/\sigma$. We have set $L/\sigma=1.00$, $\Omega_B\sigma=\Omega_A\sigma$ and $\Delta z/\sigma=1.00$.}
\label{IvsO}
\end{figure}

\begin{figure}[!htbp]
\centering
\begin{tabular}{c@{\hspace{1cm}}c}
\includegraphics[width=0.4\textwidth]{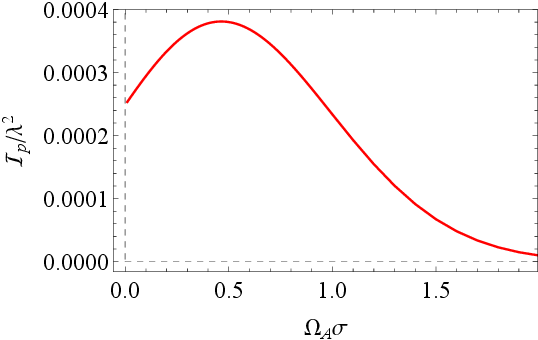} &
\includegraphics[width=0.4\textwidth]{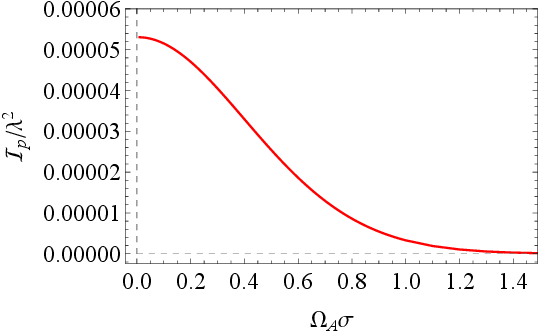} \\
\small (a) $a\sigma=0.10$ &
\small (b) $a\sigma=12.00$ \\
\end{tabular}
\caption{The mutual information $I_p/\lambda^2$ as a function of $\Omega_A\sigma$ is plotted for different $a\sigma$. We have set $L/\sigma=5.00$, $R/\sigma=0.02$, $\Omega_B\sigma=\Omega_A\sigma$ and $\Delta z/\sigma=1.00$.}
\label{IvsOL5}
\end{figure}

In Fig.~\ref{IvsO}, we present the mutual information $I_p/\lambda^2$ as a function of $\Omega_A\sigma$ for $L/\sigma=1.00$ with $\Omega_B\sigma = \Omega_A\sigma$. The mutual information exhibits similar behavior for different $\Delta z/\sigma$, and we take a fixed $\Delta z/\sigma=1.00$. When we take small $a\sigma=0.10$ or large $R/\sigma$=10.00, $I_p/\lambda^2$ decreases as $\Omega_A\sigma$ increases, which is similar to the linearly accelerated case \cite{Naeem2023}. For large $a\sigma=5.00$ and small $R/\sigma=0.02$, oscillatory behavior emerges as the energy gap increases, which is related to the periodicity of circular motion and different from the linear acceleration case \cite{Naeem2023}. The transition probability decreases monotonically with the energy gap \cite{Zhang2020,Li2025}, and the oscillation comes from $C$.

We depict the mutual information $I_p/\lambda^2$ as a function of $\Omega_A\sigma$ for $L/\sigma=5.00$ in Fig.~\ref{IvsOL5}. When we take a large $a\sigma=12.00$, $I_p/\lambda^2$ decreases as $\Omega_A\sigma$ increases. When we consider a small $a\sigma=0.10$, $I_p/\lambda^2$ initially increases, and then decreases. This behavior is similar to the linear acceleration case \cite{Naeem2023}.

\section{Conclusion}\label{sec:Conclusion}

We have studied the mutual information harvesting for two circularly accelerated detectors coupled with massless scalar fields with a reflecting boundary. As the interdetector separation increases, the mutual information exhibits oscillation for large acceleration and small radius. We first considered a small distance between the detectors and the boundary. For a small radius, when we take a small acceleration, as interdetector separation increases, the mutual information initially increases, and then decreases. For an intermediate acceleration, there may exist two peaks for the mutual information. When a large radius is considered, the mutual information first increases, then decreases. Moreover, for a fixed radius, a larger acceleration leads to a larger peak value of the mutual information. For a large distance from the boundary, when we take small acceleration or larger radius, the mutual information decreases monotonically with the increase of interdetector separation. With the increase of the distance between the detectors and the boundary, the mutual information tends to a stable value. For a not large interdetector separation or a large acceleration, the larger the energy gap difference, the smaller the harvested mutual information. There are intersections for different curves with large interdetector separation and small acceleration. For some parameters, the mutual information oscillates near the boundary, and the maximum value may be larger than the stable value. As the acceleration increases, for a small interdetector separation, the mutual information first increases, and then decreases. However, it decreases and goes to zero for large interdetector separation, and may exist intersections for different curves. For an intermediate interdetector separation, the mutual information may oscillate with the increase of acceleration. For a not large interdetector separation, as the energy gap increases, the mutual information in a small acceleration or a large radius decreases. When we take large acceleration and small radius, the mutual information first decreases, then oscillates, and finally goes to zero. For a large interdetector separation, the mutual information in small acceleration first increases, and then decreases. When we take a large acceleration, the mutual information decreases with the increase of energy gap. The oscillatory behavior appears for the combination of large acceleration and small radius, which stands for fast rotation and is related to the periodicity of circular motion. The oscillation becomes more intense near the boundary, which is related to the modes of the field as a result of the superposition of the propagating incident and reflected modes. Therefore, the oscillatory behavior can help us to understand the interaction between the rapidly rotating detectors and the quantum vacuum fluctuations. Moreover, the detectors can harvest more mutual information by rotating rapidly near the boundary.

\begin{acknowledgments}

This work was supported by the Nature Science Foundation of Shaanxi Province, China under Grant No. 2023-JC-YB-016 and the National Natural Science Foundation of China under Grant No. 11705144.

\end{acknowledgments}

\appendix
\section*{Appendix: Derivation of C}

We begin with the definition of $C$ in Eq.~(\ref{ccdef}), which can be rewritten as
\begin{align}\label{A1}
C &= \frac{\lambda^2}{\gamma_A \gamma_B} \int_{-\infty}^{\infty} dt \int_{-\infty}^{\infty} dt' \Bigg[ \chi_B(\tau_B(t)) \chi_A(\tau_A(t')) e^{i\left( \frac{\Omega_Bt}{\gamma_B} - \frac{\Omega_A t'}{\gamma_A} \right)} W(x_A(t'), x_B(t)) \notag \Bigg]\\
&=\frac{\lambda^2\sigma^2}{\gamma_A \gamma_B} \int_{-\infty}^{\infty} d\tilde{u} \int_{-\infty}^{\infty} d\tilde{s} \Bigg[ e^{-\tilde{s}^2/(2 \gamma_A^2)} e^{-\tilde{u}^2({\gamma_A^{-2}}+{\gamma_B^{-2}})/2} e^{\tilde{s} \tilde{u}/\gamma_A^2}e^{i\tilde{s}\Omega_A\sigma/\gamma_A}\notag\\
&\quad\times e^{i\tilde{u} (\Delta\Omega\sigma /\gamma_B-\Omega_A\sigma/\gamma_A +\Omega_A\sigma/\gamma_B)}W(x_A(t'), x_B(t)) \notag \Bigg]\tag{A1}
\end{align}
where we have used $\tilde{u} = t/\sigma, \tilde{s} = (t - t')/\sigma$ and $\Delta\Omega=\Omega_B-\Omega_A$. In particular, if the Wightman function is only dependent on $\tilde{s}$, Eq.~(\ref{A1}) can be further expressed as
\begin{equation}
C = \frac{\sqrt{2\pi} \lambda^2 \sigma^2}{\sqrt{\gamma_A^2 + \gamma_B^2}}
e^{-\frac{[\Omega_A\sigma(\gamma_A-\gamma_B)+\gamma_A \Delta \Omega  \sigma]{}^2}{2 \left(\gamma_A^2+\gamma_B^2\right)}} \int_{-\infty}^{\infty} d\tilde{s}
\left\lbrace e^{-\frac{\tilde{s}^2}{2(\gamma_A^2 + \gamma_B^2)}}
e^{\frac{i\sigma\tilde{s}[\gamma_B\Delta\Omega + (\gamma_A+\gamma_B)\Omega_A]  }{\gamma_A^2 + \gamma_B^2}}W(x_A(t'), x_B(t))\right\rbrace\tag{A2}
\end{equation}

\end{document}